\documentclass[a4paper,aps,prd,10pt,preprintnumbers,showpacs,twocolumn,superscriptaddress,nofootinbib,amsmath,amssymb]{revtex4-1}
\usepackage{cases}
\usepackage{graphicx}
\usepackage[utf8]{inputenc}
\usepackage[T1]{fontenc}
\usepackage{cmap}
\usepackage[normalem]{ulem} 
\def\imo{i}

\def\K{{\cal K}}

\begin{document}
\title{Black holes in Einstein-aether theory: Quasinormal modes and time-domain evolution}
\author{M. S. Churilova}\email{wwrttye@gmail.com}
\affiliation{Research Centre for Theoretical Physics and Astrophysics, Institute of Physics, Silesian University in Opava, CZ-746 01 Opava, Czech Republic}
\begin{abstract}
We propose accurate calculations of quasinormal modes of black holes in the Einstein-aether theory, which were previously considered in the literature, partially, with insufficient accuracy. We also show that the arbitrarily long-lived modes, quasiresonances, are allowed in the Einstein-aether theory as well and demonstrate that the asymptotic tails, unlike quasinormal frequencies, are indistinguishable from those in the Einstein theory.
\end{abstract}
\maketitle

\section{Introduction}
Quasinormal modes are proper oscillation frequencies of black holes, corresponding to the specific boundary conditions: purely outgoing wave at infinity and purely incoming wave at the event horizon. They do not depend on the way the perturbation was excited, but only on the black-hole parameters, which makes them a characteristic feature of the black-hole geometry, a kind of "fingerprints" of black holes. Quasinormal modes play a crucial role in the current observations of gravitational waves and, being studied during the past decades in a great number of papers, have become  an essential characteristic of a black-hole geometry \cite{Abbott:2016blz,TheLIGOScientific:2016src}. Even though there were detected signals for which the quasinormal frequencies are known with rather a small error of about a few percents \cite{Abbott:2016blz,TheLIGOScientific:2016src}, the large uncertainty in the determination of the mass and angular momentum of the black hole allows one to ascribe the same observed frequencies to a non-Kerr solution \cite{Konoplya:2016pmh} with different parameters, so that the alternative theories of gravity are not only not excluded by the current experiments, but even are not strongly constrained by observations in the gravitational  \cite{Abbott:2016blz,TheLIGOScientific:2016src} and electromagnetic \cite{Goddi:2017pfy,Akiyama:2019cqa} spectra.

Among alternative theories of gravity an interesting approach is connected with the Einstein-aether theory, which is a Lorentz-violating theory \cite{Ahmadi:2006cr,Bailey:2006fd,Kostelecky:2005ic,Rizzo:2005um,Altschul:2005za,Jacobson:2005bg,Heinicke:2005bp,Zhu:2019ura,Bhattacharjee:2018nus} endowing a spacetime with both a metric and a unit timelike vector field (aether) having a preferred time direction. It includes the Einstein relativity as a special case. Quasinormal modes of various black-hole solutions \cite{Eling,Berglund} in this theory were considered in \cite{Konoplya:2006rv,Konoplya:2006ar,Ding:2017gfw,Ding:2019tvs}, depending on the way the aether vector is chosen. For the first time quasinormal modes in the Einstein-aether theory were studied in \cite{Konoplya:2006rv,Konoplya:2006ar}, but it proved out that the black-hole solution \cite{Eling} considered in \cite{Konoplya:2006rv,Konoplya:2006ar} did not satisfy  the observed post-Newtonian behavior and, thereby, cannot describe a viable astrophysical black hole. The same is true for the so called Aether II type black-hole solution considered in \cite{Ding:2017gfw,Ding:2019tvs}. This means that those black-hole models and their spectra still may be relevant for the miniature or primordial black holes, but not for large astrophysical black holes. The Aether I type considered in \cite{Ding:2017gfw,Ding:2019tvs} is not discarded by the current experiments in the weak field regime, but, as we will show in the present paper, the data for quasinormal modes represented in \cite{Ding:2017gfw,Ding:2019tvs} suffers from the following two drawbacks:

 (i)$\,$ The lower multipoles are calculated with insufficient accuracy, so that the effect is, sometimes, smaller than the relative error.

(ii) Gravitational perturbations are reduced to the master wavelike equation in a non-self-consistent way, so that it cannot describe the gravitational spectrum even approximately.

Here we will compute quasirnomal modes for both types of aether with the help of two alternative methods: the higher-order WKB method  \cite{Mashoon,Schutz:1985zz,Iyer:1986np,Konoplya:2003ii,Matyjasek:2017psv,Konoplya:2019hlu} with the usage of Pad\'{e} approximants \cite{,Matyjasek:2017psv,Konoplya:2019hlu} and the time-domain integration  \cite{Gundlach:1993tp}. Both methods are sufficiently accurate and are in a good agreement with each other.

In addition, we will consider perturbations of a massive scalar field and show that, in a similar fashion with the Einstein theory, spectrum of massive fields in the Einstein-aether theory allows for arbitrarily long-lived quasinormal modes, called quasiresonances \cite{Ohashi:2004wr,Konoplya:2004wg,Konoplya:2006br,Konoplya:2005hr,Konoplya:2013rxa,Wu:2015fwa,Zinhailo:2018ska,Abdalla:2018ggo,Konoplya:2019hml,
Zinhailo:2019rwd,Konoplya:2017tvu,Churilova:2019qph,Churilova:2019sah}. We will show that at asymptotically late times, the quasinormal modes are suppressed by the power-law tail, which is indistinguishable from the Schwarzschild one.

The paper is organized as follows. In Sec. II, we review the essentials of the Einstein-aether theory and wavelike equations for test scalar and electromagnetic fields. Section III is devoted to the WKB and time-domain integration methods we used for finding quasinormal modes. In Sec. IV, we discuss the quasinormal modes of massless fields in the black-hole background for the Einstein-aether theory, while in Sec. V the case of a massive scalar field and existence of quasiresonances is discussed. The late-time tails are presented in Sec. VI. In Sec. VII, we give a brief remark on a wrong treatment of gravitational perturbations in a number of earlier publications. Finally, we summarize the obtained results and mention some open problems.

\section{The wave equation}
The Einstein-aether theory under consideration is described by the action \cite{Garfinkle:2011iw}
\begin{equation}\label{action}
S=\int d^4 x\sqrt{-g}\left[\frac{1}{16\pi G_{ae}}\left(\mathcal{R}+\mathcal{L}_{ae}\right)\right],
\end{equation}
where $G_{ae}$ is the aether gravitational constant, $\mathcal{L}_{ae}$ is the aether Lagrangian density,
\begin{equation}\label{Lagr}
-\mathcal{L}_{ae}=Z^{ab}_{\;\;cd}\left(\nabla_a u^c\right)\left(\nabla_b u^d\right)-\lambda\left(u^2+1\right),
\end{equation}
with
\begin{equation}\label{Z}
Z^{ab}_{\;\;cd}=c_1g^{ab}g_{cd}+c_2\delta^a_{\;c}\delta^b_{\;d}+c_3\delta^a_{\;d}\delta^b_{\;c}-c_4u^au^bg_{cd},
\end{equation}
where $c_i,\;i=1,2,3,4,$ are coupling constants of the theory. Although there are a number of severe constraints \cite{Jacobson:2008aj,Oost:2018tcv,Yagi:2013ava,Jacobson:2007fh} on the coupling constants $c_i$ (not only theoretical, but also observational), the papers \cite{Ding:2017gfw,Ding:2019tvs}, which we consider here, deal with the following theoretical ones \cite{Berglund}:
$$
0\leq c_{13}<1, \;\; 0\leq c_{14}<2,\;\; c_{13}\geq c_{14}/2,
$$
where $c_{13}=c_1+c_3$, $c_{14}=c_1+c_4$.

The metric of the spherically symmetric static Einstein-aether black-hole spacetime is given by
\begin{equation}\label{metric}
d s^2 = -f(r) d t^2 + \frac{d r^2}{f(r)}+r^2\left(\sin^2\theta d\phi^2+d\theta^2\right).
\end{equation}
The metric function has the following form:

(i)$\,$  For the first kind aether
\begin{equation}\label{metricfunctionEA1}
f(r) = 1- \frac{2M}{r} - I\left(\frac{2M}{r}\right)^4, \; I=\frac{27c_{13}}{256(1-c_{13})},
\end{equation}

(ii)$\,$ For the second kind aether
\begin{equation}\label{metricfunctionEA2}
f(r) = 1- \frac{2M}{r} - J\left(\frac{M}{r}\right)^2, \; J=\frac{c_{13}-c_{14}/2}{1-c_{13}}.
\end{equation}

Note that for the values $c_{13}=0$ (for the first kind aether) and $c_{13}=c_{14}/2$ (for the second kind aether), the metric (\ref{metric}) reduces to the Schwarzschild black-hole case.

\begin{figure}[h!]
\includegraphics[width=1\linewidth]{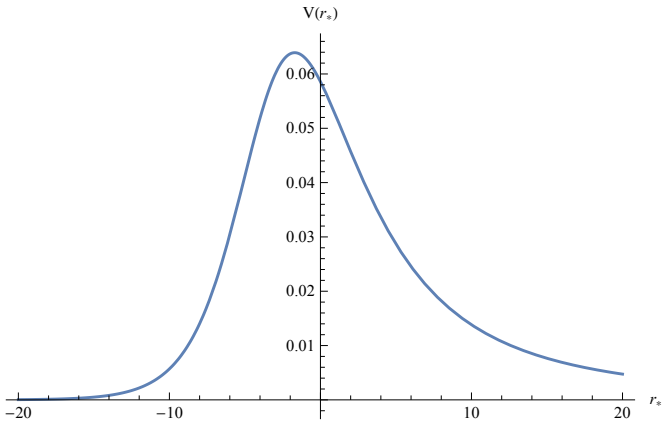}
\caption{An example of an effective potential: electromagnetic perturbations of the second kind aether black hole ($\ell=1$, $s=1$, $c=0.4$, $d=0.2$).}
\label{pot}
\end{figure}

 The general covariant equations for the test scalar $\Phi$ and electromagnetic $A_\mu$ fields have the form
\begin{equation}\label{KGg}
\frac{1}{\sqrt{-g}}\partial_\mu \left(\sqrt{-g}g^{\mu \nu}\partial_\nu\Phi\right)=0,
\end{equation}
\begin{equation}\label{EmagEq}
\frac{1}{\sqrt{-g}}\partial_\mu \left(\sqrt{-g}F^{\mu\nu}\right)=0\,,
\end{equation}
where $F_{\mu\nu}=\partial_\mu A_\nu-\partial_\nu A_\mu$.
After separation of the variables, Eqs. (\ref{KGg}) and (\ref{EmagEq}) take the following Schr\"{o}dinger-like form (see, for instance,  \cite{Konoplya:2011qq,Kokkotas:1999bd}):
\begin{equation}\label{wave-equation}
\frac{d^2\Psi_s}{dr_*^2}+\left(\omega^2-V(r)\right)\Psi_s=0,
\end{equation}
where $s=0$ corresponds to scalar field and $s=1$ to electromagnetic field and the "tortoise coordinate" $r_*$ is defined by the relation
\begin{equation}
dr_*=\frac{dr}{f(r)}.
\end{equation}
The effective potential is
\begin{equation}\label{potentialScalar}
V(r)=f(r)\left(\frac{\ell\left(\ell+1\right)}{r^2}+\frac{1-s}{r}\cdot
\frac{d\,f(r)}{dr}\right)
\end{equation}
and has the form of a potential barrier (see Fig. \ref{pot}).

\section{The methods}

\subsection{The WKB method}

The WKB method for finding quasinormal frequencies, which was first used by Schutz and Will \cite{Schutz:1985zz} (reproducing at the first order the earlier result of Mashhoon \cite{Mashoon}), grew very popular because of its effectiveness and was treated in numerous papers.

For finding quasinormal modes, we use higher-order WKB formula \cite{Mashoon,Schutz:1985zz,Iyer:1986np,Konoplya:2003ii,Matyjasek:2017psv,Konoplya:2019hlu}
\begin{eqnarray}\label{WKBformula-spherical}
\omega^2&=&V_0+A_2(\K^2)+A_4(\K^2)+A_6(\K^2)+\ldots\\\nonumber&-&\imo \K\sqrt{-2V_2}\left(1+A_3(\K^2)+A_5(\K^2)+A_7(\K^2)\ldots\right),
\end{eqnarray}
where $\K=\textrm{signRe}\left(\omega\right)\left(n+\frac{1}{2}\right)$, $n=0,1,2,3\ldots$.
The corrections $A_k(\K^2)$ of order $k$ to the first-order formula are polynomials of $\K^2$ with rational coefficients, which depend on the values $V_2, V_3\ldots$ of higher derivatives of the potential $V(r)$ in its maximum (but not on the maximum $V_0$ itself), whence it follows that the right-hand side of (\ref{WKBformula-spherical}) does not depend on $\omega$.

As the WKB method converges only asymptotically, simple increasing of the WKB formula order does not necessarily imply improving the results (see more about the asymptotic WKB regime in \cite{Hatsuda:2019eoj}). So as to increase the accuracy of the higher-order WKB formula (\ref{WKBformula-spherical}), we use Padé approximants \cite{PadeApproximation}, following Matyjasek and Opala \cite{Matyjasek:2017psv}. For the order $k$ of the WKB formula (\ref{WKBformula-spherical}), we define a polynomial $P_k(\epsilon)$ as
\begin{eqnarray}\nonumber
  P_k(\epsilon)&=&V_0+A_2(\K^2)\epsilon^2+A_4(\K^2)\epsilon^4+A_6(\K^2)\epsilon^6+\ldots\\&-&\imo \K\sqrt{-2V_2}\left(\epsilon+A_3(\K^2)\epsilon^3+A_5(\K^2)\epsilon^5\ldots\right),\label{WKBpoly}
\end{eqnarray}
whence we can obtain the squared frequency taking $\epsilon=1$,
$$\omega^2=P_k(1).$$
For the polynomial $P_k(\epsilon)$, we consider a family of the rational functions
\begin{equation}\label{WKBPade}
P_{\tilde{n}/\tilde{m}}(\epsilon)=\frac{Q_0+Q_1\epsilon+\ldots+Q_{\tilde{n}}\epsilon^{\tilde{n}}}{R_0+R_1\epsilon+\ldots+R_{\tilde{m}}\epsilon^{\tilde{m}}},
\end{equation}
called Padé approximants, with $\tilde{n}+\tilde{m}=k$, such that near $\epsilon=0$,
$$P_{\tilde{n}/\tilde{m}}(\epsilon)-P_k(\epsilon)={\cal O}\left(\epsilon^{k+1}\right).$$

It turns out that for finding fundamental mode ($n=0$) Padé approximants with $\tilde{n}\approx\tilde{m}$ usually provide the best approximation. In \cite{Matyjasek:2017psv}, $P_{6/6}(1)$ and $P_{6/7}(1)$ were compared to the sixth-order WKB formula $P_{6/0}(1)$. In \cite{Konoplya:2019hlu}, it has been observed that as a rule even $P_{3/3}(1)$ gives a more accurate value for the squared frequency than $P_{6/0}(1)$. In our case, we use sixth-order WKB expansion with appropriate Padé partition. The corresponding automatic code in \emph{Mathematica} is in open access \cite{WKB-code}.

\subsection{The time-domain integration}

If we keep in Eq. (\ref{wave-equation}) the second derivative in time instead of $\omega^2$ term, then the perturbation equations
 can be integrated at a fixed $r$ in the time domain. We use the technique of integration in the time domain developed by Gundlach  \emph{et al.} in \cite{Gundlach:1993tp}.
 We shall integrate the wavelike equation rewritten in terms of the light-cone variables $u=t-r_*$ and $v=t+r_*$. The appropriate discretization scheme is
$$
\Psi\left(N\right)=\Psi\left(W\right)+\Psi\left(E\right)-\Psi\left(S\right)
$$
\begin{equation}\label{Discretization}
-\Delta^2\frac{V\left(W\right)\Psi\left(W\right)+V\left(E\right)\Psi\left(E\right)}{8}+{\cal O}\left(\Delta^4\right)\,,
\end{equation}
where the following designations for the points were used:
$N=\left(u+\Delta,v+\Delta\right)$, $W=\left(u+\Delta,v\right)$, $E=\left(u,v+\Delta\right)$, and $S=\left(u,v\right)$. The initial data are given on the null surfaces $u=u_0$ and $v=v_0$.
To extract the values of the quasinormal modes, we shall use the Prony method (see, e.g., \cite{Berti:2007dg}) of fitting the signal by a sum of damped exponents.

\begin{center}
\begin{table*}
\begin{tabular}{p{2cm}p{3.4cm}p{1cm}p{1cm}p{1cm}p{1cm}p{3.4cm}p{1cm}p{1cm}p{1cm}p{1cm}}
\multicolumn{11}{l}{TABLE I. Fundamental quasinormal modes for the first kind aether black-hole spacetime (presented in \cite{Ding:2017gfw} (first line), obtained } \\
\multicolumn{11}{l}{here by WKB (second line) and time-domain (third line) methods).} \\[5pt]
\hline \hline
\begin{tabular}{l} \\ Parameter \\ \hline $c_{13}$ \end{tabular} & \multicolumn{5}{c}{
\begin{tabular}{p{3.4cm}p{1cm}p{1cm}p{1cm}p{1cm}} \multicolumn{5}{c}{Scalar field ($\ell=0$)} \\ \hline
$\;\;\;\;\;\;\;\;\;\;$ QNM &  \multicolumn{2}{c}{Effect \% $\;\;\;\;$} &  \multicolumn{2}{c}{Error \% $\;\;\;\;$} \\ \hline
$\;\;\;\;\;\;\;\;\;\;\;\;\;$ $\omega$ & $\delta_{\mathrm{Re}}$  & $\delta_{\mathrm{Im}}$  & $\varepsilon_{\mathrm{Re}}$  & $\varepsilon_{\mathrm{Im}}$  \end{tabular}} &
\multicolumn{5}{c}{
\begin{tabular}{p{3.4cm}p{1cm}p{1cm}p{1cm}p{1cm}} \multicolumn{5}{c}{Electromagnetic field ($\ell=1$)} \\ \hline
$\;\;\;\;\;\;\;\;\;\;$ QNM &  \multicolumn{2}{c}{Effect \% $\;\;\;\;$} &  \multicolumn{2}{c}{Error \% $\;\;\;\;$} \\ \hline
$\;\;\;\;\;\;\;\;\;\;\;\;\;$ $\omega$ & $\delta_{\mathrm{Re}}$  & $\delta_{\mathrm{Im}}$  & $\varepsilon_{\mathrm{Re}}$  & $\varepsilon_{\mathrm{Im}}$  \end{tabular}} \\[3pt]  \hline \\[-12pt]
 \begin{tabular}{l} $0$ \\[-3pt] $ $ \\[-3pt] $ $ \\[-3pt] $ $ \end{tabular} & \begin{tabular}{l} $0.104647-0.115197i$ \\[-3pt] $0.110678 - 0.104424i$ \\[-3pt] $0.109667-0.104804i$ \\[-3pt] $0.110455 - 0.104896i$ \end{tabular} &
\begin{tabular}{l} $0$ \\[-3pt] $ $ \\[-3pt] $ $ \\[-3pt] $ $ \end{tabular} & \begin{tabular}{l} $0$ \\[-3pt] $ $ \\[-3pt] $ $ \\[-3pt] $ $ \end{tabular} &
 \begin{tabular}{l} $5.4$ \\[-3pt] $ $ \\[-3pt] $ $ \\[-3pt] $ $ \end{tabular} & \begin{tabular}{l} $10$ \\[-3pt] $ $ \\[-3pt] $ $ \\[-3pt] $ $ \end{tabular} & \begin{tabular}{l} $0.245870-0.093106i$ \\[-3pt] $0.248255 - 0.092480i$ \\[-3pt] $0.248264-0.092491i$ \\[-3pt] $0.248264-0.092488i$ \end{tabular} &
  \begin{tabular}{l} $0$ \\[-3pt] $ $ \\[-3pt] $ $ \\[-3pt] $ $ \end{tabular} & \begin{tabular}{l} $0$ \\[-3pt] $ $ \\[-3pt] $ $ \\[-3pt] $ $ \end{tabular} &
 \begin{tabular}{l} $0.96$ \\[-3pt] $ $ \\[-3pt] $ $ \\[-3pt] $ $ \end{tabular} & \begin{tabular}{l} $0.68$ \\[-3pt] $ $ \\[-3pt] $ $ \\[-3pt] $ $ \end{tabular}   \\[-1pt]
 \begin{tabular}{l} $ $ \\[-13pt] $0.15$ \\[-3pt] $ $ \\[-3pt] $ $ \end{tabular} & \begin{tabular}{l} $0.103976-0.117446i$ \\[-3pt] $0.109637 - 0.105590i$ \\[-3pt] $0.108454-0.106053i$   \end{tabular} &
 \begin{tabular}{l} $0.64$ \\[-3pt] $ $ \\[-3pt] $ $  \end{tabular} & \begin{tabular}{l} $2.0$ \\[-3pt] $ $ \\[-3pt] $ $  \end{tabular} &
 \begin{tabular}{l} $5.2$ \\[-3pt] $ $ \\[-3pt] $ $  \end{tabular} & \begin{tabular}{l} $11$ \\[-3pt] $ $ \\[-3pt] $ $  \end{tabular} &  \begin{tabular}{l} $0.243928-0.094312i$ \\[-3pt] $0.246086 - 0.093413i$ \\[-3pt] $0.246093-0.093440i$  \end{tabular} &
  \begin{tabular}{l} $0.79$ \\[-3pt] $ $ \\[-3pt] $ $  \end{tabular} & \begin{tabular}{l} $1.3$ \\[-3pt] $ $ \\[-3pt] $ $  \end{tabular} &
 \begin{tabular}{l} $0.88$ \\[-3pt] $ $ \\[-3pt] $ $  \end{tabular} & \begin{tabular}{l} $0.96$ \\[-3pt] $ $ \\[-3pt] $ $  \end{tabular}  \\[-1pt]
 \begin{tabular}{l} $ $ \\[-13pt] $0.3$ \\[-3pt] $ $ \\[-3pt] $ $ \end{tabular} & \begin{tabular}{l} $0.101739-0.120032i$ \\[-3pt] $0.107641 - 0.105651i$ \\[-3pt] $0.106391-0.107705i$   \end{tabular} &
 \begin{tabular}{l} $2.8$ \\[-3pt] $ $ \\[-3pt] $ $  \end{tabular} & \begin{tabular}{l} $4.2$ \\[-3pt] $ $ \\[-3pt] $ $  \end{tabular} &
 \begin{tabular}{l} $5.5$ \\[-3pt] $ $ \\[-3pt] $ $  \end{tabular} & \begin{tabular}{l} $14$ \\[-3pt] $ $ \\[-3pt] $ $  \end{tabular} &
  \begin{tabular}{l} $0.241266-0.095728i$ \\[-3pt] $0.243201 - 0.094451i$ \\[-3pt] $0.243208-0.094523i$  \end{tabular} &
   \begin{tabular}{l} $1.9$ \\[-3pt] $ $ \\[-3pt] $ $  \end{tabular} & \begin{tabular}{l} $2.8$ \\[-3pt] $ $ \\[-3pt] $ $  \end{tabular} &
 \begin{tabular}{l} $0.8$ \\[-3pt] $ $ \\[-3pt] $ $  \end{tabular} & \begin{tabular}{l} $1.4$ \\[-3pt] $ $ \\[-3pt] $ $  \end{tabular}  \\[-1pt]
\begin{tabular}{l} $ $ \\[-13pt] $0.45$ \\[-3pt] $ $ \\[-3pt] $ $ \end{tabular} & \begin{tabular}{l} $0.096768-0.123153i$ \\[-3pt] $0.104550 - 0.107923i$ \\[-3pt] $0.104945-0.108231i$  \end{tabular} &
 \begin{tabular}{l} $7.5$ \\[-3pt] $ $ \\[-3pt] $ $  \end{tabular} & \begin{tabular}{l} $6.9$ \\[-3pt] $ $ \\[-3pt] $ $  \end{tabular} &
 \begin{tabular}{l} $7.4$ \\[-3pt] $ $ \\[-3pt] $ $  \end{tabular} & \begin{tabular}{l} $14$ \\[-3pt] $ $ \\[-3pt] $ $  \end{tabular} &
  \begin{tabular}{l} $0.237420-0.097401i$ \\[-3pt] $0.239207 - 0.095662i$ \\[-3pt] $0.239186-0.095757i$  \end{tabular} &
   \begin{tabular}{l} $3.4$ \\[-3pt] $ $ \\[-3pt] $ $  \end{tabular} & \begin{tabular}{l} $4.6$ \\[-3pt] $ $ \\[-3pt] $ $  \end{tabular} &
 \begin{tabular}{l} $0.75$ \\[-3pt] $ $ \\[-3pt] $ $  \end{tabular} & \begin{tabular}{l} $1.8$ \\[-3pt] $ $ \\[-3pt] $ $  \end{tabular}  \\[-1pt]
\begin{tabular}{l} $ $ \\[-13pt] $0.6$ \\[-3pt] $ $ \\[-3pt] $ $ \end{tabular} & \begin{tabular}{l} $0.087386-0.127661i$ \\[-3pt] $0.101186 - 0.110012i$ \\[-3pt] $0.102302-0.109778i$  \end{tabular} &
\begin{tabular}{l} $16$ \\[-3pt] $ $ \\[-3pt] $ $  \end{tabular} & \begin{tabular}{l} $11$ \\[-3pt] $ $ \\[-3pt] $ $  \end{tabular} &
 \begin{tabular}{l} $14$ \\[-3pt] $ $ \\[-3pt] $ $  \end{tabular} & \begin{tabular}{l} $16$ \\[-3pt] $ $ \\[-3pt] $ $  \end{tabular} &
  \begin{tabular}{l} $0.231411-0.099372i$ \\[-3pt] $0.233210 - 0.097008i$ \\[-3pt] $0.233154-0.097124i$  \end{tabular} &
   \begin{tabular}{l} $5.9$ \\[-3pt] $ $ \\[-3pt] $ $  \end{tabular} & \begin{tabular}{l} $6.7$ \\[-3pt] $ $ \\[-3pt] $ $  \end{tabular} &
 \begin{tabular}{l} $0.77$ \\[-3pt] $ $ \\[-3pt] $ $  \end{tabular} & \begin{tabular}{l} $2.4$ \\[-3pt] $ $ \\[-3pt] $ $  \end{tabular} \\[-1pt]
\begin{tabular}{l} $ $ \\[-13pt] $0.75$ \\[-3pt] $ $ \\[-3pt] $ $ \end{tabular} & \begin{tabular}{l} $0.072016-0.136350i$ \\[-3pt] $0.095375 - 0.112333i$ \\[-3pt] $0.097224-0.111863i$  \end{tabular} &
\begin{tabular}{l} $31$ \\[-3pt] $ $ \\[-3pt] $ $  \end{tabular} & \begin{tabular}{l} $18$ \\[-3pt] $ $ \\[-3pt] $ $  \end{tabular} &
 \begin{tabular}{l} $24$ \\[-3pt] $ $ \\[-3pt] $ $  \end{tabular} & \begin{tabular}{l} $21$ \\[-3pt] $ $ \\[-3pt] $ $  \end{tabular} &
 \begin{tabular}{l} $0.220681-0.101581i$ \\[-3pt] $0.222992 - 0.098290i$ \\[-3pt] $0.222873-0.098429i$  \end{tabular} &
 \begin{tabular}{l} $10$ \\[-3pt] $ $ \\[-3pt] $ $  \end{tabular} & \begin{tabular}{l} $9.1$ \\[-3pt] $ $ \\[-3pt] $ $  \end{tabular} &
 \begin{tabular}{l} $1.0$ \\[-3pt] $ $ \\[-3pt] $ $  \end{tabular} & \begin{tabular}{l} $3.3$ \\[-3pt] $ $ \\[-3pt] $ $  \end{tabular} \\[-1pt]
\begin{tabular}{l} $ $ \\[-13pt] $0.9$ \\[-3pt] $ $ \\[-3pt] $ $ \end{tabular} & \begin{tabular}{l} $0.051721-0.155269i$ \\[-3pt] $0.082006 - 0.114565i$ \\[-3pt] $0.084036-0.112586i$  \end{tabular} &
\begin{tabular}{l} $51$ \\[-3pt] $ $ \\[-3pt] $ $  \end{tabular} & \begin{tabular}{l} $35$ \\[-3pt] $ $ \\[-3pt] $ $  \end{tabular} &
 \begin{tabular}{l} $37$ \\[-3pt] $ $ \\[-3pt] $ $  \end{tabular} & \begin{tabular}{l} $36$ \\[-3pt] $ $ \\[-3pt] $ $  \end{tabular} &
  \begin{tabular}{l} $0.194630-0.102849i$ \\[-3pt] $0.199463 - 0.097802i$ \\[-3pt] $0.199194-0.097953i$  \end{tabular} &
  \begin{tabular}{l} $21$ \\[-3pt] $ $ \\[-3pt] $ $  \end{tabular} & \begin{tabular}{l} $10$ \\[-3pt] $ $ \\[-3pt] $ $  \end{tabular} &
 \begin{tabular}{l} $2.4$ \\[-3pt] $ $ \\[-3pt] $ $  \end{tabular} & \begin{tabular}{l} $5.2$ \\[-3pt] $ $ \\[-3pt] $ $  \end{tabular} \\[7pt]
\hline \hline
\end{tabular}
\end{table*}
\end{center}

\section{Quasinormal modes}

We considered a fundamental ($n=0$) quasinormal mode for the scalar and electromagnetic perturbations of the Einstein-aether black-hole spacetime ($M=1$). We were interested in the lower multipole numbers ($\ell=0$ for the scalar and $\ell=1$ for the electromagnetic field) because of their dominating role in the signal.

First of all, we looked at the quasinormal frequencies from \cite{Ding:2017gfw} which correspond to the Schwarzschild limit ($c_{13}=0$ for the first kind aether in Table I and $c_{13}=c_{14}/2=0.1$ for the second kind aether in Table II). As these frequencies differed from the accurate values in the second digit after the point already (for the scalar field case), we recalculated them. For this, we used two methods: the sixth-order WKB formula with Pad\'{e} approximants $P_{5/1}(1)$ and the time-domain integration. The results obtained by the both methods turned out to be in a good agreement with the accurate values for the Schwarzschild case. Therefore, we went on with our calculations, keeping the methods' parameters (such as the order of WKB series and the orders of Pad\'{e} approximants) unchanged, for the rest of the values of the parameter $c_{13}$, considered in Tables I and II.

At each step, we also found a relative effect and a relative error of the results presented in \cite{Ding:2017gfw}. A relative effect is defined as
\begin{equation}\label{EffectRe}
\delta_{\mathrm{Re}}=\frac{\left|\mathrm{Re}\, \omega_i-\mathrm{Re}\, \omega_l\right|}{\mathrm{Re}\, \omega_l}\times 100 \%,
\end{equation}
\begin{equation}\label{EffectIm}
\delta_{\mathrm{Im}}=\frac{\left|\mathrm{Im}\, \omega_i-\mathrm{Im}\, \omega_l\right|}{\mathrm{Im}\, \omega_l}\times 100 \%,
\end{equation}
where $\omega_i$ is the current value of the quasinormal mode and $\omega_l$ is the value of the quasinormal mode, which corresponds to the Schwarzschild limit. A relative error is defined by
\begin{equation}\label{ErrorRe}
\varepsilon_{\mathrm{Re}}=\frac{\left|\mathrm{Re}\, \omega_1-\mathrm{Re}\, \omega_0\right|}{\mathrm{Re}\, \omega_0}\times 100 \%,
\end{equation}
\begin{equation}\label{ErrorIm}
\varepsilon_{\mathrm{Im}}=\frac{\left|\mathrm{Im}\, \omega_1-\mathrm{Im}\, \omega_0\right|}{\mathrm{Im}\, \omega_0}\times 100 \%,
\end{equation}
where $\omega_1$ denotes the result from \cite{Ding:2017gfw} and $\omega_0$ denotes our new result at each step.

All the obtained results are presented in Tables I and II. The values of the fundamental quasinormal mode are placed one under the other: the result from \cite{Ding:2017gfw} (first line) and the results obtained here by WKB (second line) and time-domain (third line) methods. The additional fourth line (for $c_{13}=0$ in Table I and for $c_{13}=0.1$ in Table II) contains accurate values of the fundamental quasinormal mode for the Schwarzschild case. The effect and the error are calculated for the real and imaginary parts of the quasinormal frequencies obtained in \cite{Ding:2017gfw}.

\begin{center}
\begin{table*}
\begin{tabular}{p{2cm}p{3.4cm}p{1cm}p{1cm}p{1cm}p{1cm}p{3.4cm}p{1cm}p{1cm}p{1cm}p{1cm}}
\multicolumn{11}{l}{TABLE II. Fundamental quasinormal modes for the second kind aether black-hole spacetime with fixed $c_{14}=0.2$ (presented in \cite{Ding:2017gfw}} \\
\multicolumn{11}{l}{ (first line), obtained here by WKB (second line) and time-domain (third line) methods).} \\[5pt]
\hline \hline
\begin{tabular}{l} \\ Parameter \\ \hline $c_{13}$ \end{tabular} & \multicolumn{5}{c}{
\begin{tabular}{p{3.4cm}p{1cm}p{1cm}p{1cm}p{1cm}} \multicolumn{5}{c}{Scalar field ($\ell=0$)} \\ \hline
$\;\;\;\;\;\;\;\;\;\;$ QNM &  \multicolumn{2}{c}{Effect \% $\;\;\;\;$} &  \multicolumn{2}{c}{Error \% $\;\;\;\;$} \\ \hline
$\;\;\;\;\;\;\;\;\;\;\;\;\;$ $\omega$ & $\delta_{\mathrm{Re}}$  & $\delta_{\mathrm{Im}}$  & $\varepsilon_{\mathrm{Re}}$  & $\varepsilon_{\mathrm{Im}}$  \end{tabular}} &
\multicolumn{5}{c}{
\begin{tabular}{p{3.4cm}p{1cm}p{1cm}p{1cm}p{1cm}} \multicolumn{5}{c}{Electromagnetic field ($\ell=1$)} \\ \hline
$\;\;\;\;\;\;\;\;\;\;$ QNM &  \multicolumn{2}{c}{Effect \% $\;\;\;\;$} &  \multicolumn{2}{c}{Error \% $\;\;\;\;$} \\ \hline
$\;\;\;\;\;\;\;\;\;\;\;\;\;$ $\omega$ & $\delta_{\mathrm{Re}}$  & $\delta_{\mathrm{Im}}$  & $\varepsilon_{\mathrm{Re}}$  & $\varepsilon_{\mathrm{Im}}$  \end{tabular}} \\[3pt]  \hline \\[-12pt]
 \begin{tabular}{l} $0.10$ \\[-3pt] $ $ \\[-3pt] $ $ \\[-3pt] $ $ \end{tabular} & \begin{tabular}{l} $0.104647-0.115197i$ \\[-3pt] $0.110678 - 0.104424i$ \\[-3pt] $0.110366-0.104013i$ \\[-3pt] $0.110455 - 0.104896i$ \end{tabular} &
\begin{tabular}{l} $\;0$ \\[-3pt] $ $ \\[-3pt] $ $ \\[-3pt] $ $ \end{tabular} & \begin{tabular}{l} $\;0$ \\[-3pt] $ $ \\[-3pt] $ $ \\[-3pt] $ $ \end{tabular} &
 \begin{tabular}{l} $5.4$ \\[-3pt] $ $ \\[-3pt] $ $ \\[-3pt] $ $ \end{tabular} & \begin{tabular}{l} $10$ \\[-3pt] $ $ \\[-3pt] $ $ \\[-3pt] $ $ \end{tabular} & \begin{tabular}{l} $0.245870-0.093106i$ \\[-3pt] $0.248255 - 0.092480i$ \\[-3pt] $0.248259- 0.092479i$ \\[-3pt] $0.248264-0.092488i$ \end{tabular} &
  \begin{tabular}{l} $\;\;0$ \\[-3pt] $ $ \\[-3pt] $ $ \\[-3pt] $ $ \end{tabular} & \begin{tabular}{l} $\;\;0$ \\[-3pt] $ $ \\[-3pt] $ $ \\[-3pt] $ $ \end{tabular} &
 \begin{tabular}{l} $0.96$ \\[-3pt] $ $ \\[-3pt] $ $ \\[-3pt] $ $ \end{tabular} & \begin{tabular}{l} $0.68$ \\[-3pt] $ $ \\[-3pt] $ $ \\[-3pt] $ $ \end{tabular}  \\[-1pt]
\begin{tabular}{l} $ $ \\[-11pt] $0.25$ \\[-3pt] $ $ \\[-3pt] $ $ \end{tabular} & \begin{tabular}{l} $0.100755-0.114893i$ \\[-3pt] $0.107071 - 0.103500i$ \\[-3pt] $0.107006-0.103007i$   \end{tabular} &
  \begin{tabular}{l} $\;3.7$ \\[-3pt] $ $ \\[-3pt] $ $  \end{tabular} & \begin{tabular}{l} $\;0.26$ \\[-3pt] $ $ \\[-3pt] $ $  \end{tabular} &
 \begin{tabular}{l} $5.9$ \\[-3pt] $ $ \\[-3pt] $ $  \end{tabular} & \begin{tabular}{l} $11$ \\[-3pt] $ $ \\[-3pt] $ $  \end{tabular} & \begin{tabular}{l} $0.236985-0.091929i$ \\[-3pt] $0.239475 - 0.091215i$ \\[-3pt] $0.239485- 0.091214i$  \end{tabular} &
  \begin{tabular}{l} $\;\;3.6$ \\[-3pt] $ $ \\[-3pt] $ $  \end{tabular} & \begin{tabular}{l} $\;\;1.3$ \\[-3pt] $ $ \\[-3pt] $ $  \end{tabular} &
 \begin{tabular}{l} $1.0$ \\[-3pt] $ $ \\[-3pt] $ $  \end{tabular} & \begin{tabular}{l} $0.78$ \\[-3pt] $ $ \\[-3pt] $ $  \end{tabular}  \\[5pt]
\begin{tabular}{l} $ $ \\[-11pt] $0.40$ \\[-3pt] $ $ \\[-3pt] $ $ \end{tabular} & \begin{tabular}{l} $0.095828-0.114071i$ \\[-3pt] $0.102441 - 0.101957i$ \\[-3pt] $0.102778-0.101465i$   \end{tabular} &
  \begin{tabular}{l} $\;8.4$ \\[-3pt] $ $ \\[-3pt] $ $  \end{tabular} & \begin{tabular}{l} $\;0.98$ \\[-3pt] $ $ \\[-3pt] $ $  \end{tabular} &
 \begin{tabular}{l} $6.5$ \\[-3pt] $ $ \\[-3pt] $ $  \end{tabular} & \begin{tabular}{l} $12$ \\[-3pt] $ $ \\[-3pt] $ $  \end{tabular}
 & \begin{tabular}{l} $0.225711-0.090122i$ \\[-3pt] $0.228342 - 0.089303i$ \\[-3pt] $0.228354-0.089302i$  \end{tabular} &
  \begin{tabular}{l} $\;\;8.2$ \\[-3pt] $ $ \\[-3pt] $ $  \end{tabular} & \begin{tabular}{l} $\;\;3.2$ \\[-3pt] $ $ \\[-3pt] $ $  \end{tabular} &
 \begin{tabular}{l} $1.2$ \\[-3pt] $ $ \\[-3pt] $ $  \end{tabular} & \begin{tabular}{l} $0.92$ \\[-3pt] $ $ \\[-3pt] $ $  \end{tabular} \\[5pt]
\begin{tabular}{l} $ $ \\[-11pt] $0.55$ \\[-3pt] $ $ \\[-3pt] $ $ \end{tabular} & \begin{tabular}{l} $0.089374-0.112234i$ \\[-3pt] $0.096215 - 0.099294i$ \\[-3pt] $0.097314-0.099295i$  \end{tabular} &
  \begin{tabular}{l} $15$ \\[-3pt] $ $ \\[-3pt] $ $  \end{tabular} & \begin{tabular}{l} $\;2.6$ \\[-3pt] $ $ \\[-3pt] $ $  \end{tabular} &
 \begin{tabular}{l} $7.1$ \\[-3pt] $ $ \\[-3pt] $ $  \end{tabular} & \begin{tabular}{l} $13$ \\[-3pt] $ $ \\[-3pt] $ $  \end{tabular}
  & \begin{tabular}{l} $0.210705-0.087219i$ \\[-3pt] $0.213524 - 0.086273i$ \\[-3pt] $0.213538 -0.086277i$ \end{tabular} &
  \begin{tabular}{l} $\;14$ \\[-3pt] $ $ \\[-3pt] $ $  \end{tabular} & \begin{tabular}{l} $\;\;6.3$ \\[-3pt] $ $ \\[-3pt] $ $  \end{tabular} &
 \begin{tabular}{l} $1.3$ \\[-3pt] $ $ \\[-3pt] $ $  \end{tabular} & \begin{tabular}{l} $1.1$ \\[-3pt] $ $ \\[-3pt] $ $  \end{tabular}  \\[5pt]
\begin{tabular}{l} $ $ \\[-11pt] $0.70$ \\[-3pt] $ $ \\[-3pt] $ $ \end{tabular} & \begin{tabular}{l} $0.080354-0.108123i$ \\[-3pt] $0.087179 - 0.094317i$ \\[-3pt] $0.088294-0.095006i$  \end{tabular} &
  \begin{tabular}{l} $23$ \\[-3pt] $ $ \\[-3pt] $ $  \end{tabular} & \begin{tabular}{l} $\;6.1$ \\[-3pt] $ $ \\[-3pt] $ $  \end{tabular} &
 \begin{tabular}{l} $7.8$ \\[-3pt] $ $ \\[-3pt] $ $  \end{tabular} & \begin{tabular}{l} $15$ \\[-3pt] $ $ \\[-3pt] $ $  \end{tabular}
  & \begin{tabular}{l} $0.189117-0.082138i$ \\[-3pt] $0.192182 - 0.081036i$ \\[-3pt] $0.192191- 0.081046i$  \end{tabular} &
  \begin{tabular}{l} $\;23$ \\[-3pt] $ $ \\[-3pt] $ $  \end{tabular} & \begin{tabular}{l} $\;12$ \\[-3pt] $ $ \\[-3pt] $ $  \end{tabular} &
 \begin{tabular}{l} $1.6$ \\[-3pt] $ $ \\[-3pt] $ $  \end{tabular} & \begin{tabular}{l} $1.4$ \\[-3pt] $ $ \\[-3pt] $ $  \end{tabular} \\[5pt]
\begin{tabular}{l} $ $ \\[-11pt] $0.85$ \\[-3pt] $ $ \\[-3pt] $ $ \end{tabular} & \begin{tabular}{l} $0.065688-0.097327i$ \\[-3pt] $0.071753 - 0.083125i$ \\[-3pt] $0.071982-0.084297i$  \end{tabular} &
  \begin{tabular}{l} $37$ \\[-3pt] $ $ \\[-3pt] $ $  \end{tabular} & \begin{tabular}{l} $16$ \\[-3pt] $ $ \\[-3pt] $ $  \end{tabular} &
 \begin{tabular}{l} $8.5$ \\[-3pt] $ $ \\[-3pt] $ $  \end{tabular} & \begin{tabular}{l} $17$ \\[-3pt] $ $ \\[-3pt] $ $  \end{tabular}
  & \begin{tabular}{l} $0.152828-0.071437i$ \\[-3pt] $0.156164 - 0.070156i$ \\[-3pt] $0.156158- 0.070168i$  \end{tabular} &
  \begin{tabular}{l} $\;38$ \\[-3pt] $ $ \\[-3pt] $ $  \end{tabular} & \begin{tabular}{l} $\;23$ \\[-3pt] $ $ \\[-3pt] $ $  \end{tabular} &
 \begin{tabular}{l} $2.1$ \\[-3pt] $ $ \\[-3pt] $ $  \end{tabular} & \begin{tabular}{l} $1.8$ \\[-3pt] $ $ \\[-3pt] $ $  \end{tabular} \\[7pt]
\hline \hline
\end{tabular}
\end{table*}
\end{center}

\begin{center}
\begin{table*}
\begin{tabular}{p{2cm}p{4cm}p{4cm}p{2cm}cp{1cm}p{2cm}c}
\multicolumn{8}{l}{TABLE III. Fundamental quasinormal modes for the gravitational perturbations of the Reissner-Nordstr\"{o}m black-hole spacetime } \\
\multicolumn{8}{l}{ ($\ell=2$, $M=1$).} \\[2pt]
\hline \hline \\[-11pt]
\begin{tabular}{c} \\  Q \end{tabular} & \begin{tabular}{c} \\  $\;\;\;\;\;\;\;$ Reduced \end{tabular} & \begin{tabular}{c} \\  $\;\;\;\;\;\;\;$ Accurate \end{tabular} & \multicolumn{2}{c}{ \begin{tabular}{p{2cm}c}
\multicolumn{2}{c}{Effect \%}\\ \hline $\delta_{\mathrm{Re}}$  & $\delta_{\mathrm{Im}}$ \end{tabular}} & & \multicolumn{2}{c}{ \begin{tabular}{p{2cm}c}
\multicolumn{2}{c}{Error \%} \\ \hline $\varepsilon_{\mathrm{Re}}$  & $\varepsilon_{\mathrm{Im}}$ \end{tabular}} \\  \hline \\[-12pt]
0 & $0.373620 - 0.088933i$ & $0.373620 - 0.088933i$ & $\;\;\;0$ & $0 \;\;\;\;\,$ & & $\;\;\;\;0$ & $0 \;\;\;\;\,$ \\[-3pt]
0.1 & $0.374273 - 0.088986i$ & $0.373880 - 0.088962i$ & $\;\;\;0.07$ & $0.03$ & & $\;\;\;\;0.11$ & $0.03$ \\[-3pt]
0.2 & $0.376260 - 0.089142i$ & $0.374691 - 0.089046i$ & $\;\;\;0.29$ & $0.13$ & & $\;\;\;\;0.42$ & $0.11$ \\[-3pt]
0.3 & $0.379675 - 0.089399i$ & $0.376142 - 0.089185i$ & $\;\;\;0.68$ & $0.28$ & & $\;\;\;\;0.94$ & $0.24$  \\[-3pt]
0.4 & $0.384687 - 0.089748i$ & $0.378381 - 0.089371i$ & $\;\;\;1.27$ & $0.49$ & & $\;\;\;\;1.67$ & $0.42$ \\[-3pt]
0.5 & $0.391573 - 0.090164i$ & $0.381624 - 0.089584i$ & $\;\;\;2.14$ & $0.73$ & & $\;\;\;\;2.61$ & $0.65$ \\[-3pt]
0.6 & $0.400778 - 0.090592i$ & $0.386173 - 0.089781i$ & $\;\;\;3.36$ & $0.95$ & & $\;\;\;\;3.78$ & $0.90$ \\[-3pt]
0.7 & $0.413048 - 0.090900i$ & $0.392475 - 0.089872i$ & $\;\;\;5.05$ & $1.06$ & & $\;\;\;\;5.24$ & $1.14$ \\[-3pt]
0.8 & $0.429717 - 0.090796i$ & $0.401211 - 0.089621i$ & $\;\;\;7.38$ & $0.77$ & & $\;\;\;\;7.10$ & $1.31$ \\[-3pt]
0.9 & $0.453363 - 0.089298i$ & $0.413568 - 0.088311i$ & $\;\,10.69$ & $-0.70 \;\;\;$ & & $\;\;\;\;9.62$ & $1.12$ \\[-3pt]
1 & $0.490129 - 0.081661i$ & $0.431344 - 0.083440i$ & $\;\,15.45$ & $-6.18 \;\;\;$ & & $\;\;\,13.63$ & $-2.13 \;\;\;$ \\
\hline \hline
\end{tabular}
\end{table*}
\end{center}

As the values of the modes are placed one under the other, it is easy to compare them and see that the discrepancy of our results and the accurate values starts at the fourth (scalar field) or even the fifth (electromagnetic field) digit after the point, while for the results from \cite{Ding:2017gfw} these digits are respectively the second and the third. For the rest of the considered values of the parameter $c_{13}$, this tendency is kept: the deviation of the results of \cite{Ding:2017gfw} from both of our results is considerably larger than the difference between our results as such.

The error of the quasinormal frequencies obtained in \cite{Ding:2017gfw} is rather large even for the values of the parameter $c_{13}=0$, which correspond to the Schwarzschild limit ($\varepsilon_{\mathrm{Re}}=5.4 \%$ and $\varepsilon_{\mathrm{Im}}=10 \%$ for the scalar field). It can be seen that in the case of the scalar field for the values of the parameter $c_{13}$ near the Schwarzschild limit the error is greater than the effect, for the imaginary part even by an order. For larger values of $c_{13}$, which cannot promise too much accuracy, even if the error becomes less than the effect, it still remains comparable to it. Although in the case of the electromagnetic field, the situation is not so extreme, the error as yet can come to 50\% or even 110\%.

The eikonal formulas ($\ell \rightarrow \infty$) for the quasinormal modes in the Einstein-aether theory were obtained in  \cite{Churilova:2019jqx} for both types of aether.

\section{Quasiresonance}

For a massive scalar field $\Phi$ of the mass $\mu$, general covariant equation having the form
\begin{equation}\label{ScalarMass}
\frac{1}{\sqrt{-g}}\partial_\mu \left(\sqrt{-g}g^{\mu \nu}\partial_\nu\Phi\right)-\mu^2\Phi=0,
\end{equation}
there exists a phenomenon of so-called quasiresonance \cite{Konoplya:2004wg}; increasing of the field mass $\mu$ causes decreasing of the lower overtones damping rate, which means that infinitely long-lived modes appear in the spectrum.

\begin{figure*}
\resizebox{\linewidth}{!}{\includegraphics*{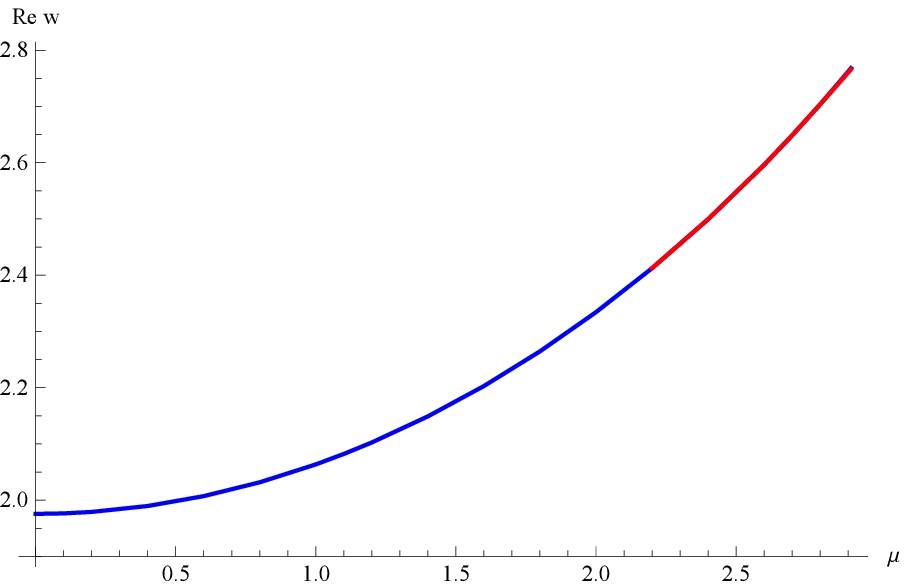}\includegraphics*{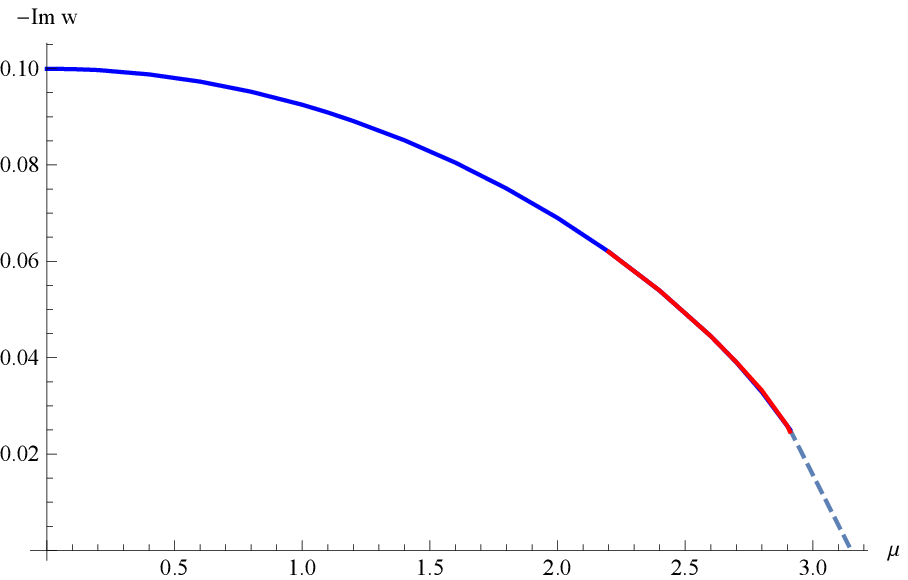}}
\caption{Real (left panel) and imaginary (right panel) parts of the fundamental quasinormal mode, calculated by the WKB method, depending on $\mu$, for the first kind aether black hole with $\ell=10$, $c_{13}=0.45$. The red part of the lines marks the values, checked by the time-domain integration.}\label{fig:QR1}
\end{figure*}

\begin{figure*}
\resizebox{\linewidth}{!}{\includegraphics*{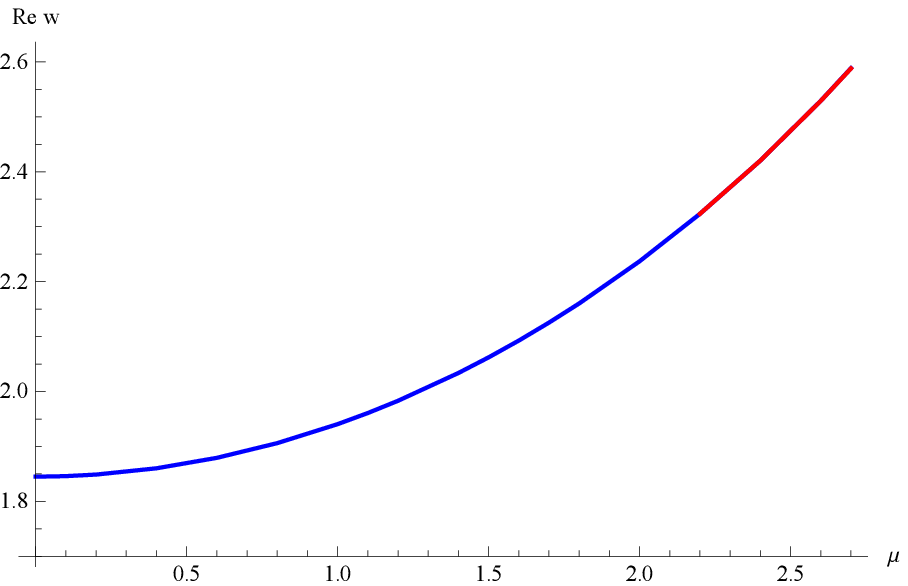}\includegraphics*{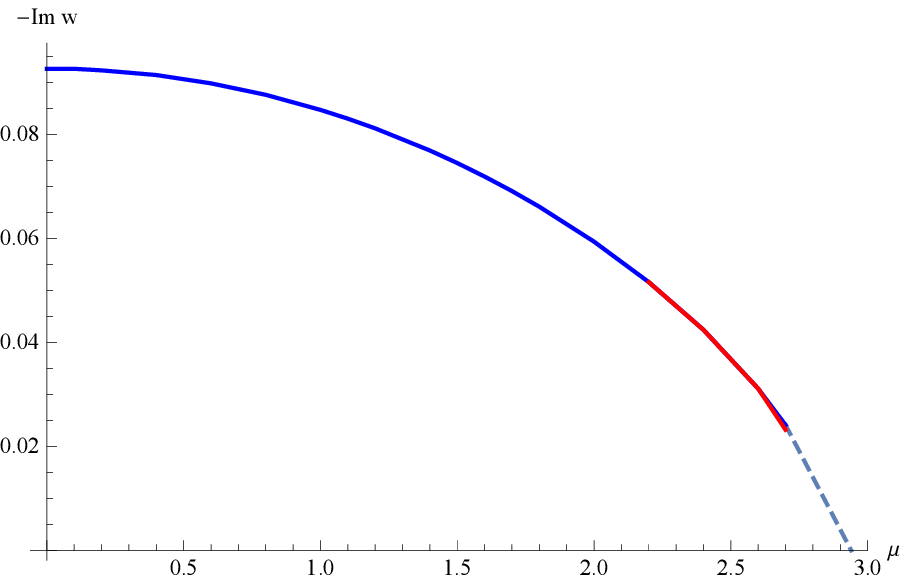}}
\caption{Real (left panel) and imaginary (right panel) parts of the fundamental quasinormal mode, calculated by the WKB method, depending on $\mu$, for the second kind aether black hole with $\ell=10$, $c_{13}=0.45$, $c_{14}=0.2$. The red part of the lines marks the values, checked by the time-domain integration.}\label{fig:QR2}
\end{figure*}

Figures \ref{fig:QR1} and \ref{fig:QR2} show dependence of the real and imaginary parts of the fundamental quasinormal mode on the mass $\mu$ of the scalar test field for the first and the second kind aether black-hole spacetime ($\ell=10$), calculated by the sixth-order WKB method with Pad\'{e} approximation. The red part of the lines marks the values of the quasinormal modes, checked by the time-domain integration (both results turned out to coincide at least up to the second digit after the point).  As WKB method works accurately when $\ell$ is much larger than $\mu M$ \cite{Konoplya:2017tvu} (although it cannot be applied in the regime of quasiresonances), the extrapolation of the WKB data can indicate the existence of quasiresonances.

\begin{figure*}
\resizebox{\linewidth}{!}{\includegraphics*{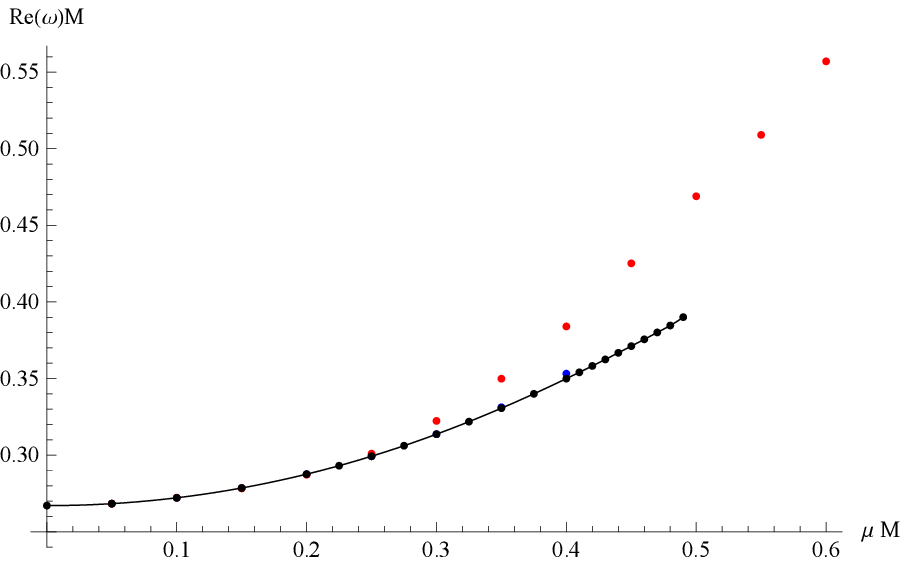}\includegraphics*{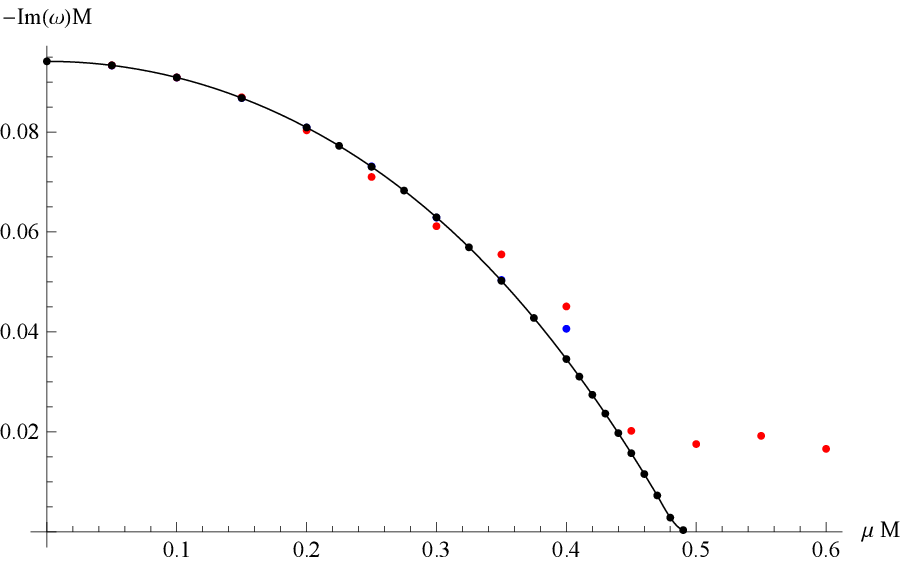}}
\caption{Real (left panel) and imaginary (right panel) parts of the fundamental quasinormal mode depending on $\mu$, for the second kind aether black hole with $\ell=1$, $c_{13}=0.45$, $c_{14}=0.2$. Blue points stand for the WKB method, red points for the time-domain integration, and black dotted line for the continued fraction method.}\label{fig:QRL1}
\end{figure*}

\begin{figure*}
\resizebox{\linewidth}{!}{\includegraphics*{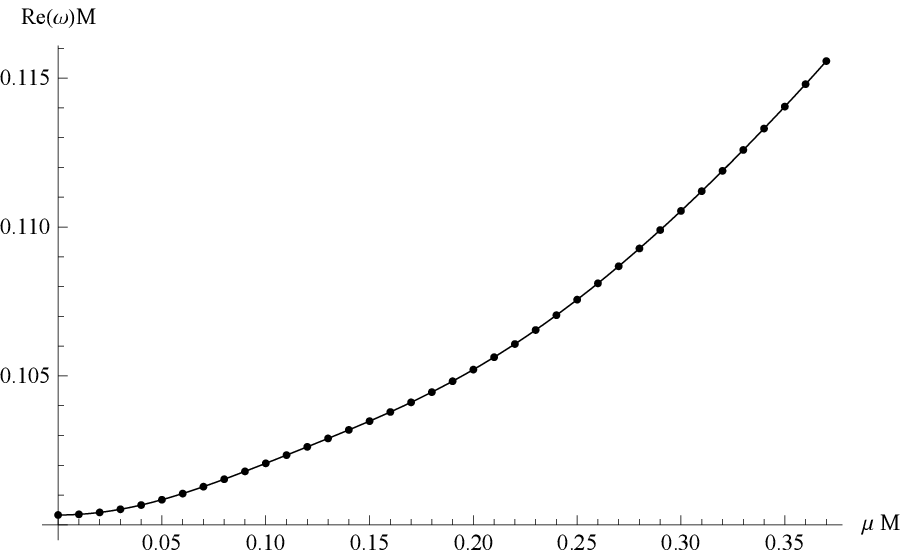}\includegraphics*{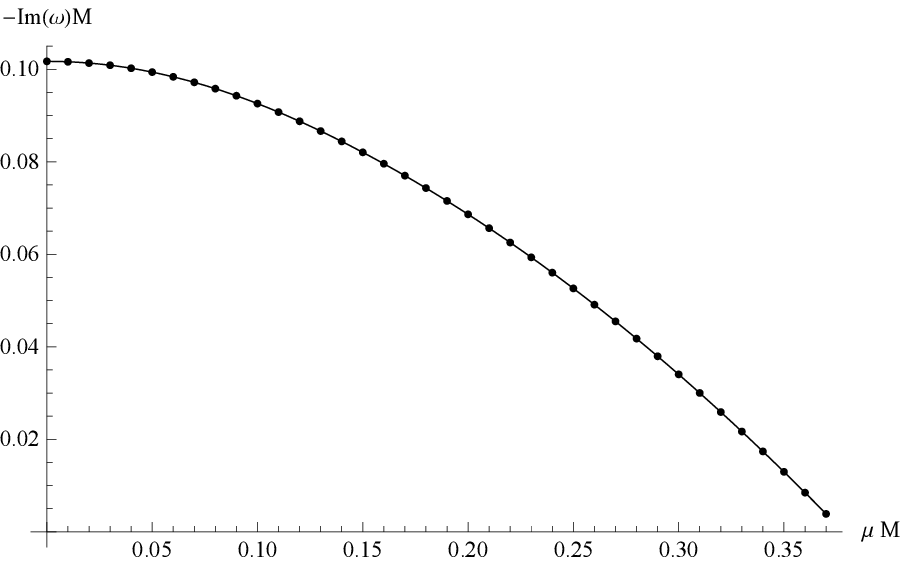}}
\caption{Real (left panel) and imaginary (right panel) parts of the fundamental quasinormal mode, calculated by the continued fraction method, depending on $\mu$, for the second kind aether black hole with $\ell=0$, $c_{13}=0.45$, $c_{14}=0.2$. }\label{fig:QRL0}
\end{figure*}

\begin{figure*}
\resizebox{\linewidth}{!}{\includegraphics*{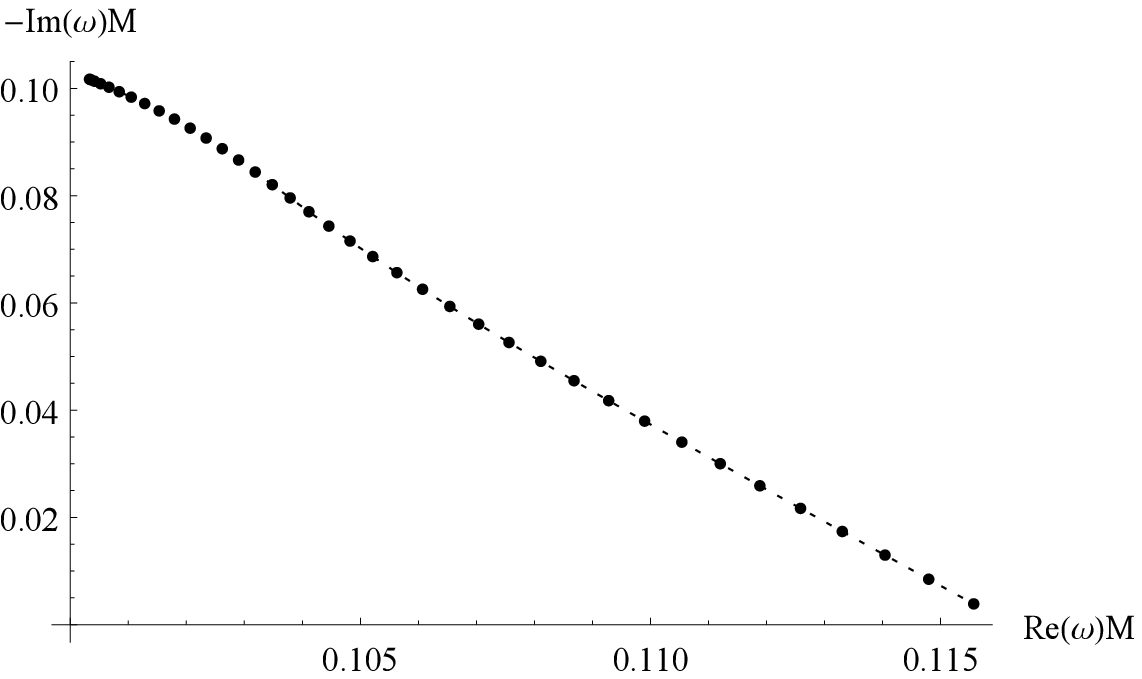}\includegraphics*{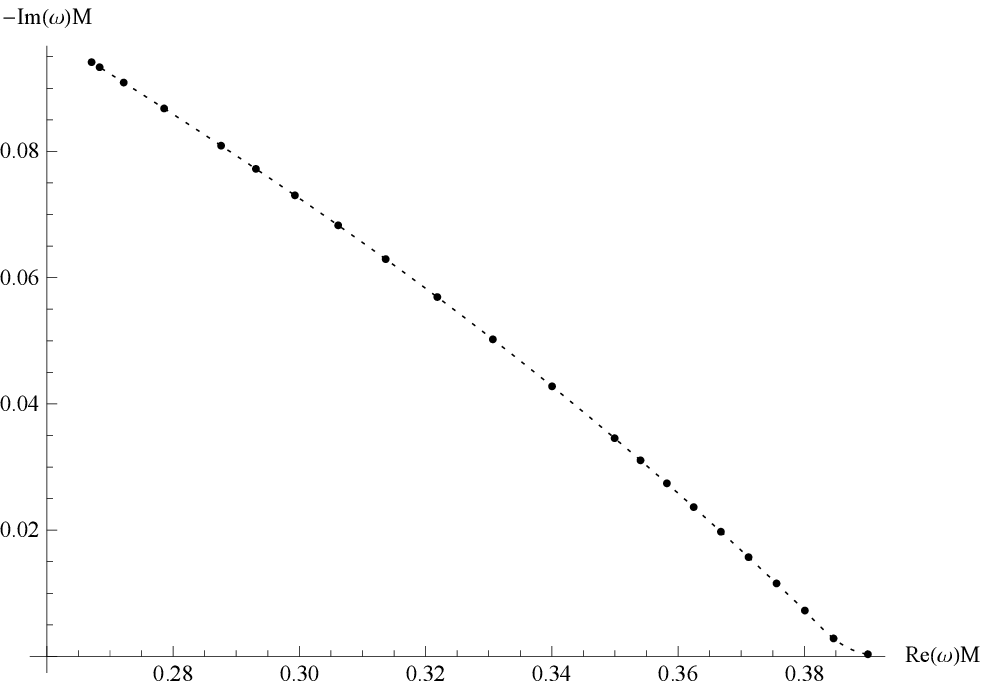}}
\caption{Dependence of the imaginary part of the fundamental quasinormal mode, calculated by the continued fraction method, on its real part for the second kind aether black hole with $\ell=0$ (left panel) and $\ell=1$ (right panel), $c_{13}=0.45$, $c_{14}=0.2$.}\label{fig:QR01}
\end{figure*}

For low multipoles, the WKB method is not always accurate, nor is even the time-domain integration (since the time-domain profile has only a few oscillations and there is a problem of extracting the value of the quasinormal mode). Therefore, for $\ell=0$ and $\ell=1$, we calculated quasinormal frequencies with the help of the continued fraction method described in \cite{Leaver,Kokkotas:2010zd}, which is convergent and gives accurate results. Figure \ref{fig:QRL1} shows the real and imaginary parts of the fundamental quasinormal mode, calculated for $\ell=1$ by all the three methods (blue points stand for the WKB method, red points for the time-domain integration, and black dotted line for the continued fraction method), depending on $\mu$. From these plots, it can be seen that the WKB method (where it is applicable) gives very close results to those obtained by the accurate continued fraction method, while the time-domain integration is not so accurate. This can be explained by appearing of the oscillating tail in the time-domain profile for the low multipole numbers. For $\ell=0$, when we cannot fully trust neither WKB method nor time-domain integration, we present in Fig. \ref{fig:QRL0} the real and imaginary parts of the fundamental quasinormal mode, calculated by the continued fraction method, depending on $\mu$. Figure \ref{fig:QR01} shows the dependence of the imaginary part of the fundamental quasinormal mode on its real part for the second kind aether black hole with $\ell=0$ and $\ell=1$.

As can be seen from the Figs. \ref{fig:QR1}--\ref{fig:QR01}, increasing of the field mass decreases the imaginary part of the quasinormal frequency, which indicates existence of the phenomenon of quasiresonance for the considered case of the massive scalar field in the Einstein-aether black-hole spacetime.

\section{Late time tails}

\begin{figure}[h!]
\includegraphics[width=1\linewidth]{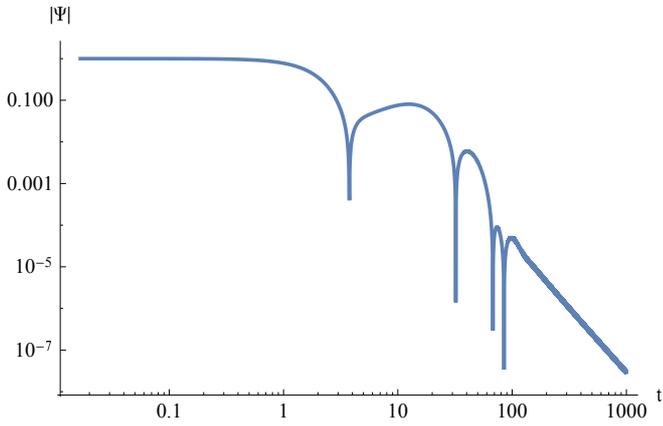}
\caption{An example of the time-domain profile: scalar perturbations of the second kind aether black hole ($\ell=0$, $s=0$, $c_{13}=0.45$, $c_{14}=0.2$).}
\label{tail}
\end{figure}

The incompleteness of the quasinormal modes set implies that at sufficiently late times the quasinormal modes are suppressed by exponential or power-law tails. Figure \ref{tail} demonstrates an example of the time-domain profile for the scalar perturbations ($s=0$, $\ell=0$) of the second kind Einstein-aether black-hole spacetime, where it can be seen that the late-times  tails for some fixed values of the black-hole parameters and $\ell=0$
$
\left|\Psi\right|\sim t^{-3}
$ are the same that those for the Schwarzschild black-hole case. Indeed, for a scalar field in the Schwarzschild background, we have the following general law:
\begin{equation}
|\Psi| \sim t^{- (2 \ell +3)}.
\end{equation}

\section{Remark on gravitational perturbations}

In a few previously published works  not only in the Einstein-aether gravitational perturbations \cite{Ding:2019tvs,Konoplya:2006ar}, but also in the Einstein-Maxwell theory \cite{Kokkotas:1993ef}, the Einstein equations,
\begin{equation}
R_{\mu \nu} - \frac{1}{2} R g_{\mu \nu} = \kappa T_{\mu \nu},
\end{equation}
were perturbed in such a way that perturbations of the right-hand side of the Einstein equations, containing the energy momentum tensor of the matter fields, were neglected.
Thus, instead of the full perturbation equations

\begin{equation}\label{full}
\delta (R_{\mu \nu} - \frac{1}{2} R g_{\mu \nu}) = \kappa \delta T_{\mu \nu},
\end{equation}
the reduced set of equations was considered,
\begin{equation}\label{reduced}
\delta (R_{\mu \nu} - \frac{1}{2} R g_{\mu \nu}) =0.
\end{equation}
This reduction was usually justified by relatively small energy content of matter fields. However, the linearized values on the right- and left-hand sides must be of the same order and cannot be ignored. There is a simple way to check whether our supposition is correct. For this, we will consider the full set of perturbation equations given by  (\ref{full}) for the Reissner-Nordstr\"{o}m spacetime as a solution of the Einstein-Maxwell equations and the corresponding reduced set given by Eq. (\ref{reduced}).
The effective potential for axial perturbations within the reduced procedure (\ref{reduced}) can be found, for example, in \cite{Ding:2019tvs}
\begin{equation}\label{potentialGravDing}
V(r)=f(r)\left(\frac{\left(\ell+2\right)\left(\ell-1\right)+2f(r)}{r^2}-\frac{1}{r}
\frac{d\,f(r)}{dr}\right),
\end{equation}
while one of the two axial potnetials for the full set perturbations of the Einstein-Maxwell field for the Reissner-Nordstr\"{o}m black hole is
$$
V(r)=\left(1-\frac{2 M}{r}+\frac{Q^2}{r^2}\right)\left(\frac{\left(\ell+2\right)\left(\ell-1\right)+2}{r^2}\right.
$$
\begin{equation}\label{potentialGravR}
\left.-\frac{\sqrt{9 M^2+4 \left(\ell+2\right)\left(\ell-1\right) Q^2}}{r^3}+\frac{16 Q^2-12 M r}{4r^4}\right).
\end{equation}
From Table III, one can see that for every value of the electric charge $Q$, the effect given by the nonzero charge in comparison with the Schwarzschild limit is smaller than or of the same order as the error due to neglecting perturbations of the energy-momentum tensor. Therefore, we conclude that such neglecting cannot be used to provide any reliable results.
Thus, the full set of perturbation equations is necessary to complement the quasinormal spectrum of the Einstein-aether black holes and to conclude about their stability.

\section{Conclusions}

In the present paper, we have shown that pervious considerations of quasinormal spectrum of black holes in the Einstein-aether theory \cite{Ding:2017gfw,Ding:2019tvs} suffer from the two main drawbacks: insufficient accuracy of reported quasinormal frequencies at lower multipoles $\ell$, such that the effect is frequently smaller than the error, and inconsistency of treatment of gravitational perturbations for which the linearization of the energy-momentum tensor cannot be neglected. Here we compute accurate quasinormal modes of massless test scalar and electromagnetic fields and, in addition, consider a massive scalar field for which we demonstrate the existence of the arbitrarily long-lived quasinormal modes called quasiresonances. We also study asymptotic tails and time-domain profiles of the Einstein-aether theory and show that at asymptotic times the tails are identical to those of the Einstein theory.

Our paper can be extended in a number of ways. First of all, we showed that consideration of the full set of perturbations equations is necessary to analyze the gravitational spectrum and, therefore, to conclude about the stability of the black hole in the Einstein-aether theory. In addition, the fermionic perturbations can be further considered in a similar way to the bosonic ones studied in this paper.

\acknowledgments{The author acknowledges Roman Konoplya for useful discussions and support  of  the  Grant No. 19-03950S of Czech Science Foundation ($GA\check{C}R$). The author also thanks Alexander Zhidenko for kind help.}

\end{document}